**Advanced colloidal nanostructures for electronic applications**

Christian Klinke, Institute of Physical Chemistry, University of Hamburg

The developments in our society in terms of computing power, sensing, and alternative energy demand new generations of building blocks for future optical and electrical applications with higher efficiencies and lower costs. Nanostructured materials such as nanoparticles, nanowires, carbon nanotubes, and graphene are considered to be among the most promising candidates for faster and less expensive electronic devices and more efficient solar cells and fuel cells. This is mostly due to their low-dimensional nature which opens new intriguing possibilities for technology. In particular, colloidal nanoparticles will find applications as fluorescence markers in medicine and biology, in flat panel displays they will be used in modern background illumination systems (Sony[TM] displays with Triluminos[TM] technology). Thin films of semiconductor nanoparticles can be used as inexpensive transistors, photo-detectors, chemical sensors, and as solar cells. Nanoparticles which show super-paramagnetic behavior should find their way into hyperthermia therapy and contrast enhancer in nuclear magnetic resonance imaging. Scientifically, low-dimensional objects also represent exceptionally interesting model systems which allow the extension of classical and the development of alternative concepts on a new field of physics with ground-breaking insights. Examples are single-electron transistors, Coulomb blockade, ballistic transport over micrometers in length, and continuously tunable optical properties. In order to produce nanomaterials with new, tailored properties it is indispensable to understand the synthesis mechanisms in detail.

*History*

Already the ancient Romans produced metallic gold nanoparticles [1]. After annealing a mixture of gold salts, soda ash, and sand they obtained transparent, deep red glass. Their method was also used for colored windows in churches of the medieval times like Notre Dame in Paris or the Cologne Cathedral. In 1659, Glauber synthesized colloidal gold in aqueous solution by reduction of gold salts with tin chloride. A similar method was described by Cassius in 1685 in his book "De Auro" [2]. Later on in 1857, Faraday delivered a synthesis and a scientific discussion of colloidal gold which was based on the reduction of gold chloride with phosphorous. These colloidal, water based suspensions showed red color, too. Nowadays, we attribute the color to the light absorption by a collective surface electron excitation, the so-called plasmon resonance [3,4]. In 1900, Ostwald produced yellow mercury oxide particles from red mercury oxide powder just by grinding [5]. This might have been already a step to quantum confined semiconductor nanocrystals. Ostwald thought that the change of surface energy is responsible for the color change. Colloidal semiconductor nanoparticles (or quantum dots) which are subject to quantum confinement have been first described by Henglein in 1982 [6]. This synthesis of cadmium sulfide nanoparticles in water, when compared to bulk CdS, showed a blue shift in the optical absorption spectrum. In the same year the brothers Efros and Efros [7] and Brus [8] explained the spectra of the nanoparticles by the quantum confinement effect which Brus calculated two years later with a handy formula, nowadays named after him [9]. A similar, but empirical formula was found for CdSe in glass by Katzschmann, Kranold and Rehfeld already in 1977 [10]. In the 1980s Ekimov, Efros, and Onushchenko synthesized CdS nanoparticles in a glass matrix and observed quantum confinement as a function of the particle size [11]. A few years later, Murray, Norris, and Bawendi introduced a new, improved method for the synthesis [12]. They injected the organometallic precursors directly into a hot coordinating solvent (Fig. 1). This allowed an instant nucleation of seeds which grew subsequently to nanoparticles. Using size-selective precipitation they obtained monodisperse colloidal cadmium chalcogenide (CdS, CdSe, CdTe) nanoparticles of tunable size. Since then, the colloidal synthesis of nanoparticles has experienced a tremendous development.

*Possible materials*

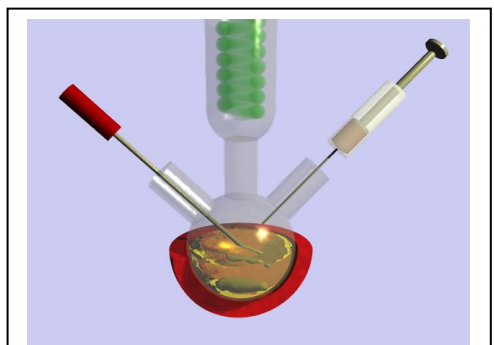

*Fig. 1: Typical setup for the synthesis of colloidal nanoparticles: a three-neck flask holds the solvent, the precursors, the ligands, and a stirring rod keeping the volume well mixed. Furthermore, there is a thermometer to control the temperature and an orifice to inject further precursors and to extract aliquots by a syringe. At the top of the setup a reflux-cooler keeps the solvent in condensed phase. The synthesis volume can be set under vacuum or nitrogen atmosphere. A heating jacket brings the synthesis on reaction temperature.*

Inorganic colloidal nanoparticles are defined by organic ligands. Such ligands are usually long-chained alkanes with a functional head group like a thiol, a carboxylic acid, a phosphonic acid, or an amine. The ligands have different tasks to fulfill: they keep the nanoparticles suspended in solution, they prevent nanoparticle aggregation, they passivate dangling bonds on the surface of the nanocrystalline particles, and last but not least they play an active role in the definition of size and shape of the nanoparticles. The use of ligands (also various ones in one synthesis) allows passivation of certain crystal facets. Most of the used ligands (such as deprotonated acids) bind strongly to metal atoms. This is in particular useful in case of compound semicondutors. In CdSe with hexagonal wurtzite crystal structure ligands such as octadecylphosphonic acid bind selectively to the cadmium sites which allows for a fast growth of the selenium-rich, poorly passivated (000-1) facet [13]. This, in turn, results in a rod-like growth of the nanoparticles. The mechanism of organic-inorganic interaction determines the size and shape of a huge number of colloidal nanoparticles. Beside precisely-defined, spherical nanoparticles with tunable diameter [12] and rod-shaped ones [13] (Fig. 2), it is possible to synthesize nanoparticles with a zig-zag shape [14], a cubic geometry [15,16], octahedral form [17], and as wires [18]. Materials composed of atoms of almost the whole periodic table can be produced. Typical examples are semiconductors with increasing bandgap like PbS, PbSe, CdS, CdSe, ZnS, and ZnO [19], and metals with increasing work function such as Ag, Au, Pd, and Pt [20]. Also doped [21] and alloyed [16] particles are feasible.

*Growth mechanisms*

Investigations on the nucleation and growth of crystals have a long history [22]. For nanoparticles it is usually described within the LaMer model [23]: The formation of nanoparticles is an interplay of surface formation energy which needs to be invested and a gain in binding energy. In a first step the complexed precursors react at elevated temperatures with their counterparts to form monomers (first units of the final material) which increases the monomer concentration. At a certain concentration nuclei are formed. They can decay again releasing surface energy. If a critical monomer concentration is reached the clusters grow fast enough to overcome a threshold in size (critical diameter). A continued growth lowers the Gibbs free energy due to a release of binding energy. From this point on the gain in binding energy is larger than the loss due to surface formation. Thus, the nanoparticles continue to grow; no decay takes plays anymore. The growth decreases the monomer concentration. If it falls short of the critical concentration no further nucleation takes place. As a consequence, in order to obtain a monodisperse diameter distribution the period in which the monomer concentration exceeds the critical concentration (and stable nucleation takes place) must be as short as possible. The growth goes on until most of the monomer is consumed. If the monomer concentration is comparatively high, those first stages are kinetically controlled. The shape can differ from the thermo-dynamically most favored one due to a fast kinetic growth as described above for CdSe rods. At later stages, thermodynamic aspects become more important. Ripening sets in: At elevated temperatures surface atoms can rearrange to form thermodynamically more favored shapes (e.g. spheres or pyramids). This process can also be induced by organic ripening agents. For example, halogenated organic compounds like 1,2-dichloroethane can trigger the transformation of rod-like CdSe nanoparticles to hexagonal pyramids [24]. Released halides bind as charged, atomic ligands preferably to cadmium-rich facets which leads to the thermodynamically favored pyramidal shape. The less stable the halogen alkane is the faster the ripening takes place. The introduction of halogenated hydrocarbons represents a new degree-of-freedom in nanoparticle shape control. Without the use of triggering molecules inter-particle ripening is possible, for example the so-called Ostwald ripening: In order to minimize the total surface (energy) small particles dissolve and larger ones grow in size [25]. Organic ligands on the nanoparticles' surface can prevent ripening effects to some extent and stabilize facets and the particle shape.

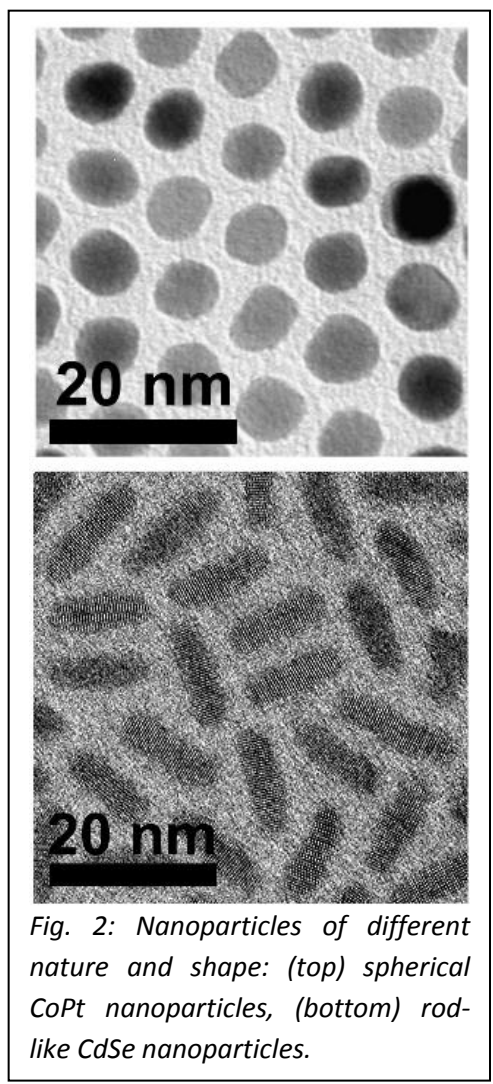

*Fig. 2: Nanoparticles of different nature and shape: (top) spherical CoPt nanoparticles, (bottom) rod-like CdSe nanoparticles.*

Another effect that can take place, especially for poorly passivated particles, is "oriented attachment": In the later stages of growth nanoparticles agglomerate and merge in order to minimize the total surface. Usually, this happens *via* the most reactive crystal facets of the nanoparticles [26] or by dipole-dipole interaction [27].

Often nanoparticles suffer from incomplete passivation of the dangling bonds by ligands. This is due to a poor interaction of a facet with the ligands or due to steric hindrance. Poorly passivated semiconducting nanoparticles show weak fluorescence properties due to the presence of resulting trap states. The passivation can be improved by a post-synthetic ligand exchange for better-passivating ones. An even better approach is to grow an inorganic shell around the original particles which has a similar lattice constant but a higher bandgap (e.g. CdSe core with a CdS shell). Depending on the band-alignment both, the optically excited electron and hole, are confined in the core (Type I configuration) or will be separated (Type II configuration). To improve the optical properties further yet another shell can be grown around the initial core-shell structure (e.g. CdSe/CdS with a further ZnS shell). This approach was shown to be successful not only for spherical particles but also for rods [28] and stars [17].

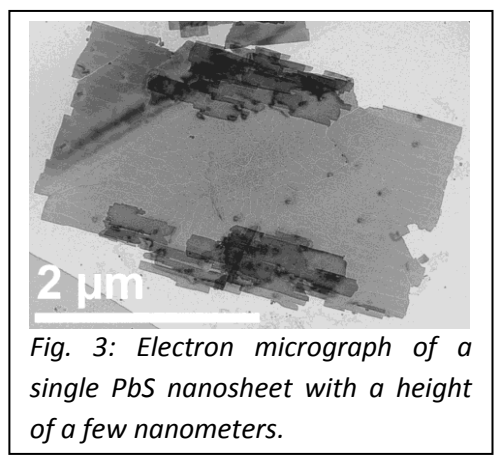

*Fig. 3: Electron micrograph of a single PbS nanosheet with a height of a few nanometers.*

A recent development is the synthesis of colloidal nanosheets (aka nanoplatelets). In 2009, Hyeon *et al.* introduced the synthesis of CdSe platelets with lateral extensions of about 100 nm and a height of only 1.4 nm [29]. In some systems, such as in CdTe platelets, it is possible also to tune the height by the synthetic conditions [30]. Those two systems, CdSe and CdTe, possess a wurtzite crystal structures and the platelets grow most likely continuously by reaction with precursors. In contrast, large PbS nanosheets, with a cubic crystal symmetry grow by oriented attachment [31] (Fig. 3). Small, readily formed PbS nanoparticles merge over the {110} facets to form sheets which have lateral extensions of a few micrometers and which are only a few nanometers in height. At a first glance, it is difficult to understand why a cubic material like PbS grows in only two dimensions. Long-chained alkane ligands tend to self-assemble and to form two-dimensional structures by themselves (self-assembled monolayers, SAM). The assumption is that the self-assembly of long-chained ligands assists the two-dimensional aggregation of nanoparticles which eventually merge to complete two-dimensional monocrystals. Two-dimensional colloids are not limited to semiconductors. It is also possible to synthesize metal platelets, e.g. made of gold [32] and palladium [33].

*Hybrid nanoparticles*

To extend the tool box for the synthesis of functional nanostructures seed nanoparticles can be introduced. The seeds possess already a size which is larger than the critical radius. Thus, other materials can readily grow on top of the seed materials. Using low melting point materials like bismuth nanoparticles, it is possible to solve precursors in those seed particles. An oversaturation in the molten nanoparticles then leads to a growth in one direction. In this way, the growth of long nanowires becomes feasible [18]. Catalytically active metal nanoparticles can also be generated during the synthesis [34]: At moderate temperatures thermal decomposition of trimethylindium leads to small metallic indium droplets. They act as catalyst for the decomposition of the present ligand trioctylphosphine to available phosphorous. Together these two compounds form needle-like structures of the semiconductor material indium phosphide InP. Thus, the resulting structure is a semiconductor InP needle with a metal In head. Electronically speaking, the interface In/InP forms a well-conducting Ohmic contact; whereas gold electrodes in contact to InP leads to a hurdle for electrons, a so-called Schottky barrier. As a result, a gold-contacted In/InP nanoneedle can function as a ready-made Schottky diode and transistor (Fig. 4).

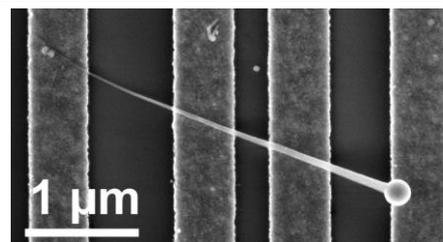

*Fig. 4: Gold electrodes (vertical bars) contact an InP nanoneedle with a spherical In head. The device works as ready-made Schottky diode and transistor.*

In another approach further materials can be combined with pre-synthesized structures in a similar way like what has been described for core-shell particles. Preferentially, the secondary material grows on well-defined sites on the primary one. The idea is that the properties of the individual structures should either complement one another or new properties emerge. For example, soft magnetic $CoFe_2O_4$ nanoparticles were decorated with silver particles [35]. In the optical absorption spectra the combined dumbbell particles show a surface plasmon resonance stemming from the silver part. The polarization plane of a monochromatic beam going through a suspension of $CoFe_2O_4$ nanoparticles rotates by a certain angle depending on the magnitude of the applied magnetic field (Faraday effect). Interestingly, the presence of silver can stabilize (no crossover due to the dielectric contribution of silver) and enhance the rotation signal compared to pure $CoFe_2O_4$ particles.

The synthesis of guest materials on host nanoparticles can be led quite precisely. The addition of gold chloride complexed with ammonium bromide and oleylamine to a suspension of CdSe nanopyramids in toluene leads to an instable gold shell around the structures [36]. The introduction of a reducing agent yields then stable metallic gold clusters at the most reactive sites of the CdSe nanopyramids, the vertices. Depending on the amount of added gold precursor the cluster size can be as small as 1.4 ± 0.3 nm (corresponding to about 50 atoms). The gold clusters grow on the surface of the CdSe nanopyramids with defined interfaces (Fig. 5). Such hybrid structures could be useful in photocatalysis and water splitting.

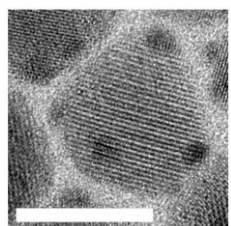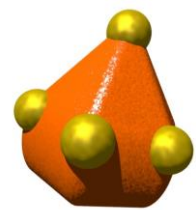

*Fig. 5: CdSe nanopyramid covered with gold clusters at the vertices (Left: Electron microscopy image, the scale bar corresponds to 10 nm; Right: Model of the structure).*

*Composites*

Another interest lies in the attachment of nanoparticles to one-dimensional systems like carbon nanotubes. It is anticipated that they can be beneficial in third-generation solar cells [37]: Incident light gets absorbed by semiconductor nanoparticles generating electron-hole pairs. Due to the energy alignment the charge carriers get separated and the carbon nanotubes lead one type of charge carrier (e.g. the electrons) efficiently to one contact. The opposite type of charge carrier can be received by a wrapped around conductive polymer. Furthermore, metallic particles can serve as catalysts to create branches on nanotubes [38], and magnetic nanoparticles attached to nanotubes yield highly anisotropic magnetic materials [39]. In previous approaches, carbon nanotubes were washed in acids to generate defects in the lattice structure. Subsequently, a few nanoparticles could be linked to the tubes covalently [40]. The drawback of this method is that it modifies the nanotubes in terms of conductance, mechanical properties, and optical behavior. A generic, mild method to attach all kinds of nanoparticles to carbon nanotubes, singlewalled and multiwalled ones, with a high coverage has been established just in the recent years [41] (Fig. 6). The investigations on such composites show that the interaction between nanotubes and nanoparticles is of electrostatic nature and can be understood as a ligand exchange where nanotubes stabilize the nanoparticles [16].

The synergy between semiconductor nanoparticles and carbon nanotubes has been shown for zinc oxide nanoparticles on doublewall nanotubes [42]. Single-wall tubes, usually used for carbon nanotube field-effect transistors, tend to bundle in organic solvents which prevents the attachment of nanoparticles to individual nanotubes. For this reason doublewall nanotubes had been chosen. They can be dispersed individually in organic solvent. The resulting composites were spread on Si/SiO$_2$ wafers and gold electrodes were deposited on top of the individual composite structures. The resulting devices function as field-effect transistors. Photoelectrical transport measurements showed that electron-hole pairs generated by exposure to light can be separated in such structures. Due to the adsorption of oxygen on ZnO such devices are not only sensitive to light but also able to detect the oxygen particle pressure.

*Physical properties*

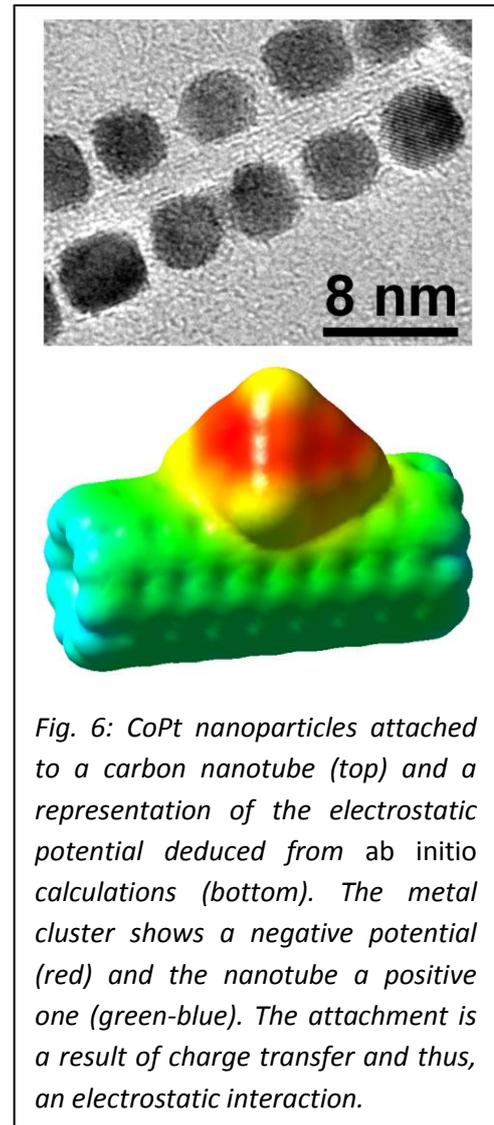

*Fig. 6: CoPt nanoparticles attached to a carbon nanotube (top) and a representation of the electrostatic potential deduced from* ab initio *calculations (bottom). The metal cluster shows a negative potential (red) and the nanotube a positive one (green-blue). The attachment is a result of charge transfer and thus, an electrostatic interaction.*

The electronic confinement of charge carriers in nanoparticles has a decisive impact on their optical properties. The most obvious effect is that the absorption and fluorescence wavelength shifts with decreasing nanoparticle volume to shorter wavelengths. In a simple approach this can be explained by the quantum mechanical textbook model of a particle in a box. Depending on the effective mass of the electrons (holes) and the size of the extensions the bulk conduction (valence) band shifts up (down) resulting in an increased bandgap. Further, the Coulomb interaction of electron and hole (the exciton) in confined space should be considered. For spherical particles the effective bandgap can be described by the Brus equation [9]: $\Delta E_{g,eff} = \Delta E_{g,bulk} + h^2(1/m_e^*+1/m_h^*)/8R^2 - 1.786 \cdot e^2/4\pi\varepsilon_o\varepsilon_r^\infty R$ with the effective bandgap $\Delta E_{g,eff}$, the bulk bandgap $\Delta E_{g,bulk}$, the Planck constant $h$, the effective mass for electrons (holes) in the material $m_e^*$ ($m_h^*$), the elementary charge $e$, the vacuum permittivity $\varepsilon_o$, the relative high-frequency permittivity $\varepsilon_r^\infty$, and the particle radius $R$. The factor 1.786 stems from the overlap integral of the electron and hole wavefunction in the exciton. Of course the actual optical properties depend also on the solvent, the type of stabilizing ligand, *etc.*

Anyhow, this model has been applied successfully in a lot of cases. By modern optical microscopy methods it is possible to identify individual fluorescing nanoparticles. In case of individual nanoparticles the effect of "blinking" occurs [43]: times of fluorescence are followed of dark periods and *vice versa*. This is explained by a trapping of one type of charge carrier of the optically excited exciton in surface states. The remaining opposite type of charge carrier can then be excited to higher energies by the absorption of a following photon. This process prevents fluorescence as long as the first charge carrier is trapped. After relaxation of the first charge carrier to the ground state regular fluorescence is possible again. Another sign for individual particles is their spectroscopic line shape. Since the nanoparticles can be considered as individual oscillators absorbing and emitting photons the line shape is the one of a damped oscillator which can be fitted with a Lorenz distribution. In an ensemble measurement plenty of such oscillators add to a broadened line shape which can be fitted with a Gaussian bell curve.

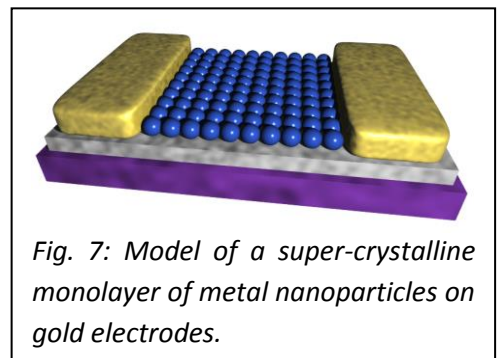

*Fig. 7: Model of a super-crystalline monolayer of metal nanoparticles on gold electrodes.*

The size of the particles of a few nanometers has also a decisive impact on their electrical properties. Charging the particles with an electron requires an energy of $E_C = e^2/C$ with the elementary charge $e$ and the capacity of the nanoparticle $C$. Due to their size they possess a very small electrical capacity and thus, a comparatively large Coulomb charging energy. Thus, electrons can only be added in discrete steps. This classical effect is called Coulomb blockade. Each further charging of the nanoparticle requires again the same energy which results in a "Coulomb energy ladder". In order to investigate the charge transport over a nanoparticle it must not be contacted directly. Otherwise the nanoparticles would be part of the electrode and possess a much larger capacity. This can be accomplished by contacting the nanoparticles *via* tunnel barriers. In order to pass a current through a nanoparticle the electrons must tunnel form a first electrode onto the nanoparticle overcoming the Coulomb charging energy and tunnel to a second electrode. A third electrode located further away does not make direct contact to the nanoparticles (tunneling processes are not possible either); but it can shift the Coulomb energy levels of the nanoparticle up or down electrostatically.

Technologically interesting are films of nanoparticles. They can serve as inexpensive transistors [44] or solar cells [45]. A simple method to produce such films is spin-coating or dip-coating. For more precisely defined structures the well-known Langmuir-Blodgett technique can be exploited [46] (Fig. 7). Usually, it is used to establish monolayers of amphiphilic molecules, but it can also be used to produce well-ordered monolayers of colloidal nanoparticles [47]. Deposited on gold electrodes, the electronic transport through films of metal nanoparticles shows that they behave almost ohmically at room temperature [48]. At lower temperatures and moderate bias voltages the electrons do not possess enough energy to overcome the Coulomb energy. Thus, the conductance is zero. Higher bias voltages lower the Coulomb levels and electrons can tunnel from particle to particle through the whole film. The magnitude of the conductance is a matter of particle size (proportional to the single-particle capacity) and interparticle distance. The distance between the particles (tunnel barrier width) is determined by the length of their organic ligands. Thermal annealing of the films transforms a small fraction of the ligands into amorphous, conductive carbon [49]. The annealing time and temperature determine the amount of carbon between the particles, bridging the tunnel barriers. As a result, the conductance through the films increases. This tunability of the film conductance is useful e.g. for sensors: The trade-off between base current and sensitivity can be optimized for the corresponding application.

As mentioned before, the position of the Coulomb energy levels can be manipulated by an external electrical field. This holds good not only for individual particles but also for films – at least for monolayers (In multilayered films the metallic particles would screen the electrical field). If a dielectric layer is deposited on the metal nanoparticle monolayer a third, gate electrode can be defined on top of it by electron-beam lithography. The application of a gate voltage can then modulate the current through the film [50]. This is how a transistor works. In this case, a voltage sweep leads to oscillations of the current – in accordance with the Coulomb energy ladder. Larger nanoparticles and thicker dielectric layers lead to reduced oscillations. At temperatures above the Coulomb blockage regime the oscillations are completely lifted.

Although the tunnel barriers proved to be beneficial in metal nanoparticle films, since they defined the capacity of the particles, in thin, semiconducting films they are disadvantageous since they decrease the electrical performance of the devices. Besides the annealing of the films in order to carbonize the ligands, it is also possible to use thioltetrazole ligands in the synthesis. During thermal annealing, those molecules decompose mainly to smaller, gaseous fragments. The nanoparticles move closer together and the conductance of the films increases [51]. Another approach is to replace the original long-chained aliphatic ligands by short inorganic ligands [44], by semiconducting ones [52], or by charged, atomic ligands such as $Cl^-$ [45].

In all aforementioned approaches to reduce the height or the width of the effective tunnel barriers nonetheless grain boundaries remain. This is of advantage in thermoelectric materials which generate a voltage based on temperature differences. Efficient thermoelectrica rely on unhampered electrical transport while the thermal conductivity is reduced compared to continuous, macroscopic materials. The thermal conductivity can be lowered by introduction of interfaces in the material which lead the electrical current almost lossless but block the propagation of lattice vibrations, the so-called phonons, which contribute to the heat transport. Materials made of bismuth telluride nanoparticles proved to be promising candidates for future thermoelectric applications [53]. The removal of the ligands on the nanoparticles leads to a precipitation of the particles. A subsequent moderate sintering yields a compact, almost continuous material with well-defined grains of the size of the nanoparticles. The material shows highly reduced thermal conductivity and the same electric conductivity as bulk bismuth telluride.

For transistors it is expedient to avoid tunnel barriers from the beginning on. In this respect, sheet-like colloidal materials are interesting alternatives. They are suspended in solvents and thus, solution processable. Additionally, they naturally possess no tunnel barriers in the plane, and their electrical properties are tunable by their height. Their superior properties could be demonstrated in individual PbS nanosheets which were used as field-effect transistors [54]. They show high on/off current ratios, high field-effect mobilities, and strong response to exposure to light.

Nowadays the synthesis of colloidal nanostructures is very versatile. They can be based on a large amount of materials; they can possess plenty sizes and shapes, and they can be combined to more complex structures. This gives them an enormous variety of properties and makes them hot candidates for future applications, especially in optics and electronics.

**References**

[1] M. C. Daniel, D. Astruc, Chem. Rev. 104 (2004) 293.
[2] L. B. Hunt, Gold Bulletin 9 (1976) 134.
[3] M. Faraday, Philos. Trans. R. Soc. 147 (1857) 145.
[4] W. Ostwald, Z. Chem. Ind. Koll. 4 (1909) 5.
[5] W. Ostwald, Z. Phys. Chem. 34 (1900) 495.
[6] Z. Alfassi, D. Bahnemann, A. Henglein, J. Phys. Chem. 86 (1982) 4656.
[7] A. L. Efros, Al. Efros, Sov. Phys. Semicond. 16 (1982) 772.
[8] R. Rossetti, L. Brus, J. Phys. Chem. 86 (1982) 4470; R. Rossetti, S. Nakahara, L. Brus, J. Chem. Phys. 79 (1983) 1086.
[9] L. E. Brus, J. Chem. Phys. 80 (1984) 4403.
[10] R. Katzschmann, R. Kranold, A. Rehfeld, Phys. Stat. Sol. A 40 (1977) K161.
[11] A. I. Ekimov, Al. Efros, A. A. Onushchenko, Sol. State Commun. 56 (1985) 921.
[12] C. B. Murray, D. J. Norris, M. G. Bawendi, J. Am. Chem. Soc. 115 (1993) 8706.
[13] Z. A. Peng, X. Peng, J. Am. Chem. Soc. 124 (2002) 3343; B. H. Juarez, C. Klinke, A. Kornowski, H. Weller, Nano Lett. 7 (2007) 3564.
[14] K. S. Cho, D. V. Talapin , W. Gaschler, C. B. Murray, J. Am. Chem. Soc., 127 (2005) 7140.
[15] D. V. Talapin, H. Yu , E. V. Shevchenko, A. Lobo , C. B. Murray, J. Phys. Chem. C 111 (2007) 14049.
[16] B. Ritz, H. Heller, A. Myalitsin, A. Kornowski, F. J. Martin-Martinez, S. Melchor, J. A. Dobado, B. H. Juarez, H. Weller, C. Klinke, ACS Nano 4 (2010) 2438.
[17] M. Scheele, N. Oeschler, I. Veremchuk, S. O. Peters, A. Littig, A. Kornowski, C. Klinke, H. Weller, ACS Nano 5 (2011) 8541.
[18] Z. Li, A. Kornowski, A. Myalitsin, A. Mews, Small 4 (2008) 1698.
[19] Y. W. Jun, J.-S. Choi, J. Cheon, Angew. Chem. Int. Ed. 45 (2006) 3414.
[20] A. R. Tao, S. Habas, P. Yang, Small 4 (2008) 310.
[21] K. Riwotzki, M. Haase, J. Phys. Chem. B 102 (1998) 10129.
[22] R. Becker, W. Döring, Annalen der Physik 416 (1935) 719.
[23] V. K. La Mer, R. H. Dinegar, J. Am. Chem. Soc. 72 (1950) 4847; V. K. La Mer, Ind. Eng. Chem. 44 (1952) 1270.
[24] M. Meyns, F. Iacono, C. Palencia, J. Geweke, M. D. Coderch, U. E. A. Fittschen, J. M. Gallego, R. Otero, B. H. Juarez, C. Klinke, Chem. Mater. 26 (2014) 1813.
[25] W. Ostwald, Z. Phys. Chem. 22 (1897) 289.
[26] C. Pacholski, A. Kornowski, H. Weller, Angew. Chem. Int. Ed. 41 (2002) 1188.
[27] J. Polleux, N. Pinna, M. Antonietti, M. Niederberger, Adv. Mater. 16 (2004) 436.
[28] M. Saba, S. Minniberger, F. Quochi, J. Roither, M. Marceddu, A. Gocalinska, M. V. Kovalenko, D. V. Talapin, W. Heiss, A. Mura, G. Bongiovanni, Adv. Mater. 21 (2009) 4942; D. Dorfs, A. Salant, I. Popov, U. Banin, Small 4 (2008) 1319.
[29] J. S. Son, X.-D. Wen, J. Joo, J. Chae, S.-I. Baek, K. Park, J. H. Kim, K. An, J. H. Yu, S. G. Kwon, S.-H. Choi, Z. Wang, Y.-W. Kim, Y. Kuk, R. Hoffmann, T. Hyeon, Angew. Chem. Int. Ed. 48 (2009) 6861.

[30] S. Pedetti, B. Nadal, E. Lhuillier, B. Mahler, C. Bouet, B. Abecassis, X. Xu, B. Dubertret, Chem. Mater. 25 (2013) 2455.
[31] C. Schliehe, B. H. Juarez, M. Pelletier, S. Jander, D. Greshnykh, M. Nagel, A. Meyer, S. Förster, A. Kornowski, C. Klinke, H. Weller, Science 329 (2010) 550.
[32] X. Huang, S. Li, Y. Huang, S. Wu, X. Zhou, S. Li, C. L. Gan, F. Boey, C. A. Mirkin, H. Zhang, Nature Comm. 2 (2011) 292.
[33] X. Huang, S. Tang, X. Mu, Y. Dai, G. Chen, Z. Zhou, F. Ruan, Z. Yang, N. Zheng, Nature Nanotech. 6 (2011) 28.
[34] T. Strupeit, C. Klinke, A. Kornowski, H. Weller, ACS Nano 3 (2009) 668.
[35] Y. Li, Q. Zhang, A. V. Nurmikko, S. Sun, Nano Lett. 5 (2005) 1689.
[36] M. Meyns, N. G. Bastus, Y. Cai, A. Kornowski, B. H. Juarez, H. Weller, C. Klinke, J. Mater. Chem. 20 (2010) 10602.
[37] W. Feng, Y. Feng, Z. Wu, A. Fujii, M. Ozaki, K. Yoshino, J. Phys.: Condens. Matter 17 (2005) 4361; H. Borchert, F. Witt, A. Chanaewa, F. Werner, J. Dorn, T. Dufaux, M. Kruszynska, A. Jandke, M. Holtig, T. Alfere, J. Bottcher, C. Gimmler, C. Klinke, M. Burghard, A. Mews, H. Weller, J. Parisi, J. Phys. Chem. C 116 (2012) 412.
[38] C. Klinke, E. Delvigne, J. V. Barth, K. Kern, J. Phys. Chem. B 109 (2005) 21677.
[39] V. Salgueirino-Maceira, M. A. Correa-Duarte, M. Banobre-Lopez, M. Grzelczak, M. Farle, L. M. Liz-Marzan, J. Rivas, Adv. Funct. Mater. 18 (2008) 616.
[40] S. Banerjee, S. S. Wong, Chem. Commun. (2004) 1866.
[41] B. H. Juarez, C. Klinke, A. Kornowski, H. Weller, Nano Lett. 7 (2007) 3564; B. H. Juarez, M. Meyns, A. Chanaewa, Y. Cai, C. Klinke, H. Weller, J. Am. Chem. Soc. 130 (2008) 15282; A. B. Hungria, B. H. Juarez, C. Klinke, H. Weller, P. A. Midgley, Nano Research 1 (2008) 89.
[42] A. Chanaewa, B. H. Juarez, H. Weller, C. Klinke, Nanoscale 4 (2012) 251.
[43] X. Wang, X. Ren, K. Kahen, M. A. Hahn, M. Rajeswaran, S. Maccagnano-Zacher, J. Silcox, G. E. Cragg, Alexander L. Efros, T. D. Krauss, Nature 459 (2009) 686.
[44] A. Nag, M. V. Kovalenko, J.-S. Lee, W. Liu, B. Spokoyny, D. V. Talapin, J. Am. Chem. Soc. 133 (2011) 10612.
[45] A. H. Ip, S. M. Thon, S. Hoogland, O. Voznyy, D. Zhitomirsky, R. Debnath, L. Levina, L. R. Rollny, G. H. Carey, A. Fischer, K. W. Kemp, I. J. Kramer, Z. Ning, A. J. Labelle, K. W. Chou, A. Amassian, E. H. Sargent, Nature Nanotech. 7 (2012) 577.
[46] I. Langmuir, J. Am. Chem. Soc. 39, 1848 (1917); I. Langmuir, K. B. Blodgett, Kolloid-Zeitschrift 73, 257 (1935); K. B. Blodgett, J. Am. Chem. Soc. 57, 1007 (1935).
[47] V. Aleksandrovic, D. Greshnykh, I. Randjelovic, A. Frömsdorf, A. Kornowski, S. V. Roth, C. Klinke, H. Weller, ACS Nano 2 (2008) 1123.
[48] D. Greshnykh, A. Frömsdorf, H. Weller, C. Klinke, Nano Lett. 9 (2009) 473.
[49] Y. Cai, D. Wolfkuhler, A. Myalitsin, J. Perlich, A. Meyer, C. Klinke, ACS Nano 5 (2011) 67.
[50] Y. Cai, J. Michels, J. Bachmann, C. Klinke, J. Appl. Phys. 114 (2013) 034311.
[51] J. Lauth, J. Marbach, A. Meyer, S. Dogan, C. Klinke, A. Kornowski, H. Weller, Adv. Funct. Mater. 24 (2014) 1081.
[52] M. Scheele, D. Hanifi, D. Zherebetskyy, S. T. Chourou, S. Axnanda, B. J. Rancatore, K. Thorkelsson, T. Xu, Z. Liu, L. W. Wang, Y. Liu, A. P. Alivisatos, ACS Nano (ASAP).
[53] M. Scheele, N. Oeschler, K. Meier, A. Kornowski, C. Klinke, H. Weller, Adv. Func. Mater. 19 (2009) 3476.
[54] S. Dogan, T. Bielewicz, Y. Cai, C. Klinke, Appl. Phys. Lett. 101 (2012) 073102.